# Refractive displacement of the radio-emission footprint of inclined air showers simulated with CoREAS

Felix Schlüter[1,2,a], Marvin Gottowik[3,b], Tim Huege[1,4], Julian Rautenberg[3]

[1] Karlsruher Institut für Technologie, Institut für Kernphysik, Karlsruhe, Germany
[2] Universidad Nacional de San Martín, Instituto de Tecnologías en Detección y Astropartículas, Buenos Aires, Argentina
[3] Bergische Universität Wuppertal, Wuppertal, Germany
[4] Vrije Universiteit Brussel, Astrophysical Institute, Brussels, Belgium

**Abstract** The footprint of radio emission from extensive air showers is known to exhibit asymmetries due to the superposition of geomagnetic and charge-excess radiation. For inclined air showers a geometric early-late effect disturbs the signal distribution further. Correcting CoREAS simulations for these asymmetries reveals an additional disturbance in the signal distribution of highly inclined showers in atmospheres with a realistic refractive index profile. This additional apparent asymmetry in fact arises from a systematic displacement of the radio-emission footprint with respect to the Monte-Carlo shower impact point on the ground. We find a displacement of ∼ 1500 m in the ground plane for showers with a zenith angle of 85°, illustrating that the effect is relevant in practical applications. A model describing this displacement by refraction in the atmosphere based on Snell's law yields good agreement with our observations from CoREAS simulations. We thus conclude that the displacement is caused by refraction in the atmosphere.

**Contents**

[a] e-mail: felix.schlueter@kit.edu (corresponding author)
[b] e-mail: gottowik@uni-wuppertal.de

## 1 Introduction

Radio detection of inclined air showers is a powerful technique for the detection of ultra-high energy cosmic rays. It has been demonstrated that the detection of these particles is possible with sparse antenna arrays [1]. The radio emission, which originates from the electromagnetic component of an air shower, experiences almost no attenuation while propagating through the atmosphere and hence provides an accurate and precise energy estimator [2]. Combined with a measurement of the muonic shower component, e.g. by a particle detector, measurements of inclined air showers also provide a high mass-composition sensitivity [3], which will be in particular a target of the large-scale radio detector of the upcoming upgrade of the Pierre Auger Observatory [4]. Therefore, inclined air showers have a high relevance and a detailed understanding of the signal distribution of their radio emission is crucial to accurately reconstruct the cosmic-ray properties of interest.

The radio signal distribution is affected by a strong asymmetry arising from the superposition of the geomagnetic and charge-excess emission caused by their individual polarization patterns [5]. For inclined showers with a zenith angle larger than 60° an additional early-late asymmetry becomes relevant [6].

In [7] a previously unknown apparent asymmetry in the radio-emission footprint of inclined CoREAS simulations was observed which was presumed to be related to refraction of the radio emission in the atmosphere. In this paper we explain and resolve this apparent asymmetry with a systematic offset of the radio-emission footprint with respect to the Monte-Carlo (MC) air shower impact point, i.e., the intersection between the MC shower axis and a ground plane. We express this offset as a displacement of the center of symmetry of the dominating geomagnetic radio emission from the Monte-Carlo impact point. We develop a method to determine the radio symmetry center without implying detailed knowledge of the lateral distribution function (LDF) of the radio-emission footprint. Furthermore, we present a model successfully describing the radio symmetry center displacement by the refraction of electromagnetic waves propagating through a refractive atmosphere based on Snell's law.

In [8] the influence of the refractive index on the radio-emission footprint and thus the reconstructed depth of the shower maximum $X_{max}$ was studied. An important correlation of the reconstructed $X_{max}$ with the refractivity at the shower maximum was found. Focusing on vertical showers, effects on signal propagation were, however, neglected.

The effect presented here is crucial in order to gain a more detailed understanding of the signal distribution of radio emission from inclined air showers. Implications arise for the modeling and reconstruction of air showers as well as the interpretation of the reconstructed geometries. Our investigation mainly refers to the frequency band of the radio emission from 30 to 80 MHz. This frequency band is used by most current-generation large scale radio detector arrays [9–11] as well as the radio detector of the upgrade Pierre Auger Observatory [4]. However, many next generation radio experiments [12,13] aim to cover higher frequencies and a larger band, e.g. 50 to 200 MHz for the proposed GRAND experiment [12]. In Sect. 3.4 we will therefore address the comparability of our results with this frequency band.

This article is structured as follows: In the following section we show that the additional apparent asymmetry in the lateral distribution of inclined showers found in [7] can be explained with a displaced radio symmetry center. In Sect. 3 we present a method to determine the symmetry center of the lateral distribution of the radio emission. Using a set of CoREAS showers, we establish a systematic displacement of the radio symmetry center with respect to the MC impact point. Furthermore, we investigate differences in the Cherenkov radius for geomagnetic and charge-excess emissions. In Sect. 4 we present a model based on Snell's law that successfully describes the displacement, and we discuss the treatment of the propagation of radio emission in CoREAS. Finally, we draw our conclusions in Sect. 5.

## 2 Apparent asymmetry in the signal distribution of air-shower radio emission

For vertical air showers, several LDF models exist which take the interference between geomagnetic and charge-excess emission into account [14,15]. It is assumed that the geomagnetic and charge-excess radiation, independently, have a rotationally symmetric emission pattern around the shower axis. However, [2] reports relative deviations in the energy fluence of the sub-dominant charge-excess contribution of $\lesssim 20\%$. In inclined air showers the radio emission above the shower axis travels over longer distances between source and observer than below axis. The intensity of the expanding radio emission scales with this geometrical distance. Consequently below the shower axis, i.e., early in the shower, observers at ground measure a higher signal intensity than observers late in the shower and thus an early-late asymmetry is imprinted in the radio-emission footprint. A geometrical correction consisting of a "projection into the shower plane" assuming a point source of the radio emission located at the shower maximum $X_{max}$ can resolve this asymmetry within 2% [16].

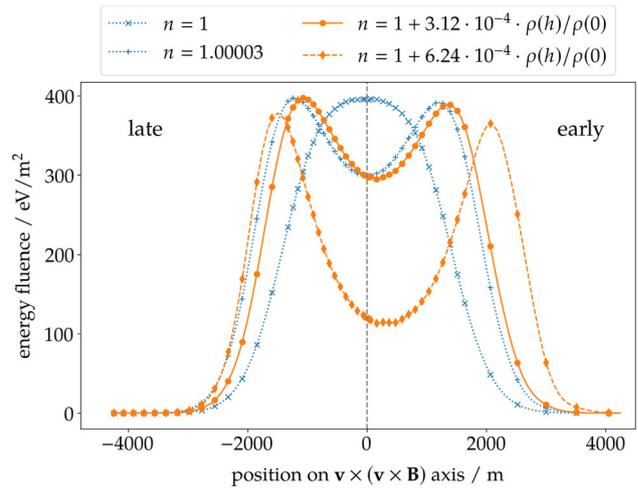

**Fig. 1** Lateral distribution of the radio emission of a 85° air shower along the positive and negative $\mathbf{v} \times (\mathbf{v} \times \mathbf{B})$ axes with respect to the MC impact point. The energy fluence is expected to be symmetric on both axes. This is fulfilled for a constant refractive index $n$ (blue lines), independent of its exact value. If the refractive index changes with height (orange lines), the LDF is not symmetric with respect to the MC shower axis. On closer look, a displacement of the symmetry axis rather than an asymmetry is observed. Figure updated from [7]

In addition to these nowadays well-known asymmetries, a further *apparent* asymmetry was observed in [7]. To further investigate this finding, we simulated one air shower with a zenith angle of 85° arriving from South for four different atmospheric refractivity profiles using the CoREAS code [17]. In Fig. 1 the lateral distribution of the radio signal along the $\mathbf{v} \times (\mathbf{v} \times \mathbf{B})$ axis is shown in terms of the energy fluence $f$ with the unit eV/m$^2$. In our notation the vector $\mathbf{v}$ points in the direction of movement of the shower particles, i.e., the direction of the primary particle and the magnetic field vector $\mathbf{B}$ points to the north and upwards direction with an inclination of $\sim 36°$. Thus observers along the positive (negative) $\mathbf{v} \times (\mathbf{v} \times \mathbf{B})$ axis are early (late). Along this axis the geomagnetic and charge-excess contributions are decoupled by their polarisation [2] and after a correction for geometrical early-late effects no asymmetry is expected. If the refractive index is set to a constant value of $n \equiv 1$ or $n \equiv 1.00003$ (approximately the value of $n(h_{\mathrm{max}})$ at the shower maximum for an air shower with a zenith angle of 85°) throughout the atmosphere, the signal distribution in the shower plane is symmetric along the positive and negative $\mathbf{v} \times (\mathbf{v} \times \mathbf{B})$ axes with respect to the MC shower axis. The exact value of the refractive index is not important for symmetry, but changes the shape of the LDF. With a changing refractive index following the density gradient in the atmosphere, an apparent asymmetry is observed, the LDF is not symmetric anymore with respect to the MC shower axis. This is enhanced when doubling the refractivity $n - 1$ throughout the atmosphere. It seems that for these simulations the symmetry axis is displaced from the MC shower axis in the direction of the positive $\mathbf{v} \times (\mathbf{v} \times \mathbf{B})$ axis rather than exhibiting an additional asymmetry. In the next section we will show that this displacement of the radio-emission footprint with respect to the MC impact point is eminent in all simulations. Thereby we will illustrate the nature of this displacement which hints towards refraction of the radio emission. In Sect. 4 we will compare this displacement with a model calculation employing refraction of the radio emission in the atmosphere.

## 3 Displacement of the radio-emission footprints

We now analyse the apparent asymmetry introduced by the refractive index profile of the atmosphere in detail using a set of 4308 inclined showers simulated with CORSIKA [18] (pre-release version of the package V7.7000) and its CoREAS extension [17].

The simulations contain proton and iron primaries with energies $18.4 \leq \log_{10}(E/\mathrm{eV}) \leq 20.2$ in $\log_{10}(E/\mathrm{eV}) = 0.2$ steps, zenith angles from 65° to 85° with a step size of 2.5°, and 8 equidistantly spaced azimuth angles $\phi$, i.e. coming from geomagnetic East ($\phi = 0°$), North-East ($\phi = 45°$), North ($\phi = 90°$), etc. For each simulation the simulated pulses are located on a star-shaped grid along the $\mathbf{v} \times \mathbf{B}$ and $\mathbf{v} \times (\mathbf{v} \times \mathbf{B})$ axes and their bisections, i.e. on concentric rings in the shower plane projected onto the ground. In the following we will refer to the pulses with the same polar angle in the shower plane as "arms" of the star-shaped grid. The spacing is denser close to the shower axis to sample the energy fluence distribution in the 30 to 80 MHz band precisely, and sparser outside, cf. Figs. 1, 2.

The simulation settings match the ambient conditions of the Pierre Auger Observatory [19] which will measure the radio emission from inclined air showers after the upcoming deployment of a large-scale radio detector [4]. The ground plane is set to an observation level of 1400 m above sea level. The atmospheric model fits the average conditions of Malargüe (location of the Pierre Auger Observatory) in October [20]. The refractivity at sea level is set to $n_0 - 1 = 3.12 \times 10^{-4}$. We use QGSJetII-04 [21] and UrQMD [22] as high- and low-energy hadronic interaction models and set a thinning level of $5 \times 10^{-6}$ with optimized weight limitation [23].

### 3.1 Fitting the Cherenkov ring

So far there is no established LDF for horizontal air showers that can be used to fit the radio symmetry center. New models are currently being developed, e.g. [16], but the results are still being validated. We therefore use a purely geometrical approach that exploits the Cherenkov compression of the radio signal for the estimation. At a certain angle, the Cherenkov angle, a large fraction of the emission, released during the complete shower evolution, arrives at the same time at ground and enhances the signal strength on a ring around the shower axis, the so-called Cherenkov ring. We here estimate the radio symmetry center by a fit of a ring to this feature and define the center of this ring as our radio symmetry center. This approach also yields an estimator for the radius of the Cherenkov ring. Its radius depends on the geometrical distance to the source region and on the refractivity in this region [8,24]. Since geomagnetic and charge-excess emission were found to originate from slightly different regions in the atmosphere [2] it is expected that both emission contributions exhibit an independent Cherenkov ring. Hence, in the following we will describe the radio-emission footprint in terms of the geomagnetic energy fluence $f_{\mathrm{geo}}$ and charge-excess energy fluence $f_{\mathrm{ce}}$ separately. Both can be calculated from the energy fluence in the $\mathbf{v} \times \mathbf{B}$ and $\mathbf{v} \times (\mathbf{v} \times \mathbf{B})$ polarisation $f_{\mathbf{v} \times \mathbf{B}}$, $f_{\mathbf{v} \times (\mathbf{v} \times \mathbf{B})}$ for a given symmetry center position via (derived from [15])

$$f_{\mathrm{geo}} = \left( \sqrt{f_{\mathbf{v} \times \mathbf{B}}} - \frac{\cos \Phi}{|\sin \Phi|} \cdot \sqrt{f_{\mathbf{v} \times (\mathbf{v} \times \mathbf{B})}} \right)^2 \quad (1)$$

$$f_{\mathrm{ce}} = \frac{1}{\sin^2 \Phi} \cdot f_{\mathbf{v} \times (\mathbf{v} \times \mathbf{B})} \quad (2)$$

where $\Phi$ denotes the polar angle of the pulse position in the shower plane with respect to the positive $\mathbf{v} \times \mathbf{B}$ axis counting counterclockwise. Equations (1) and (2) are only valid if the pulses of geomagnetic and charge-excess emission arrive at ground almost simultaneously, i.e., without a significant phase shift giving rise to a circularly polarized signal component. Following the analysis presented in [25] we find an average time delay of $\lesssim 1\,\mathrm{ns}$ for pulses within and around the Cherenkov ring. Given pulse widths of tens of nanoseconds this delay is negligible independent of the relative strength of the charge-excess emission or the considered frequency band.[1] Therefore, Eqs. (1) and (2) are valid within the scope of our analysis.

---

[1] For pulses in the hundreds of MHz or GHz regime this might no be true anymore as such pulses can be considerably shorter.

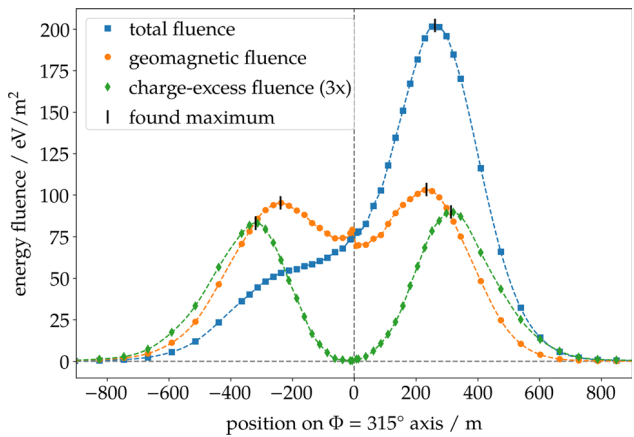

**Fig. 2** Comparison of the geomagnetic, charge-excess and total energy fluence (cf. Eqs. 1, 2) for a shower coming from North with a zenith angle of 65°. This corresponds to a geomagnetic angle with $\sin\alpha \approx 0.19$. The charge-excess contribution is multiplied by a factor of 3. Observers are shown on an axis with $\Phi = 315°$, negative values corresponds to the $\Phi = 135°$ axis. The found maxima are marked by black vertical lines on the LDF. For the geomagnetic energy fluence, a non-physical behavior can be seen close to the axis. This is an artifact of using the MC impact point as the radio symmetry center in the calculation

In Fig. 2 an example shower with a small geomagnetic angle $\alpha$ (angle between the magnetic field axis and shower axis) is shown. For such showers with a weak geomagnetic emission the interference between geomagnetic and charge-excess emission impacts the position of the maximal fluence and can even completely suppress the Cherenkov ring in the negative $\mathbf{v} \times \mathbf{B}$ half-plane. Note that since the amplitudes of the electric field traces are interfering, the resulting asymmetry in energy fluence, e.g., the squared sum of the amplitudes, is accentuated. For the total fluence no ring can be estimated for signals with $\Phi = 135°$. In contrast, the geomagnetic and charge-excess energy fluences individually exhibit a clear maximum. Note that for Fig. 2 the MC impact point was used for the calculation of the geomagnetic energy fluence which does not describe the true radio symmetry center as we will see later. The LDF described by $f_{ce}$ exhibits a broader Cherenkov ring than $f_{geo}$ which motivates to describe both features independently. Thus, in the following we will fit the Cherenkov ring to the individual footprints of the geomagnetic or charge-excess emission contributions. However, following Eqs. (1) and (2), the calculation of $f_{geo}$ and $f_{ce}$ depends on the location of the radio symmetry center via the polar angle $\Phi$. Thus it is not possible to find the locations of the Cherenkov ring on each arm of the star-shaped grid independently. Therefore, we fit the ring in an iterative process recalculating $f_{geo}$ and $f_{ce}$ in each step of the minimization. The Cherenkov ring is described by the position along each arm of the star-shaped grid, for which the fluence is maximal. These positions are found using a cubic spline interpolation along each arm of the star-shaped grid. For the minimiza-tion process we employ a least squares method with equal weights for each ring position. The calculation of $f_{geo}$ and $f_{ce}$ following Eqs. (1) and (2) becomes nonphysical for small values of $\sin \Phi$, hence the $\mathbf{v} \times \mathbf{B}$ axis is excluded. We note that pulses along this axis will not remain at $\Phi = 0°$ respectively $\Phi = 180°$ for the fitted center position and therefore the above equations could provide reasonable energy fluences using the true radio symmetry center position. However, a varying number of data points during the fit could result in a bias.

An example fit to the geomagnetic emission for an event with a zenith angle of 85° coming from North-West is shown in Fig. 3. The impact of the underlying interpolation function and the spacing of interpolated points used to find the maxima on each arm is found to be negligible for the obtained results. The displacement between the radio symmetry center and MC impact point is estimated to be 125 m in the shower plane. This is a small effect compared to the fitted Cherenkov ring radius of 1198 m, however, due to the high inclination this corresponds to a displacement of 1428 m on ground. The maximal difference between the Cherenkov radii, found on the individual arms, amounts to 40 m. The fit yields an uncertainty of the symmetry center displacement in the shower plane of 21 m.

Having two different Cherenkov rings, i.e., the ring in $f_{geo}$ and $f_{ce}$, encoded in the total signal of $f_{\mathbf{v} \times \mathbf{B}}$ and $f_{\mathbf{v} \times (\mathbf{v} \times \mathbf{B})}$ with a similar strength, i.e., for showers with a small $\sin \alpha$, makes it challenging to disentangle them. Hence, in the following we will fit the Cherenkov ring and evaluate a displacement of the emission footprint of the dominating geomagnetic emission and only for showers with a larger geomagnetic angle. For a subset of showers with a small geomagnetic angle we will compare the Cherenkov radii independently for both emission contributions.

### 3.2 Investigation of the radio symmetry center displacement for showers with a large $\sin \alpha$

For showers with a large geomagnetic angle, fulfilling $\sin \alpha > 0.25$, we determine the radio symmetry center displacement by a fit to the Cherenkov ring. This condition excludes 120 showers coming from North with a zenith angle below 70°, which we will discuss separately in Sect. 3.3. In total 4185 fits yield an accurate result and are analyzed in the following.

We interpret our results as function of the geometric distance $d_{max}$ from the MC impact point to $X_{max}$, given by

$$X_{ground} - X_{max} = \int_0^{d_{max}} \rho(\ell) \, d\ell \quad (3)$$

The atmospheric slant depth measured along the shower axis of the ground plane is denoted by $X_{ground}$, $\rho(\ell)$ denotes the atmospheric density at the distance $\ell$ along the MC

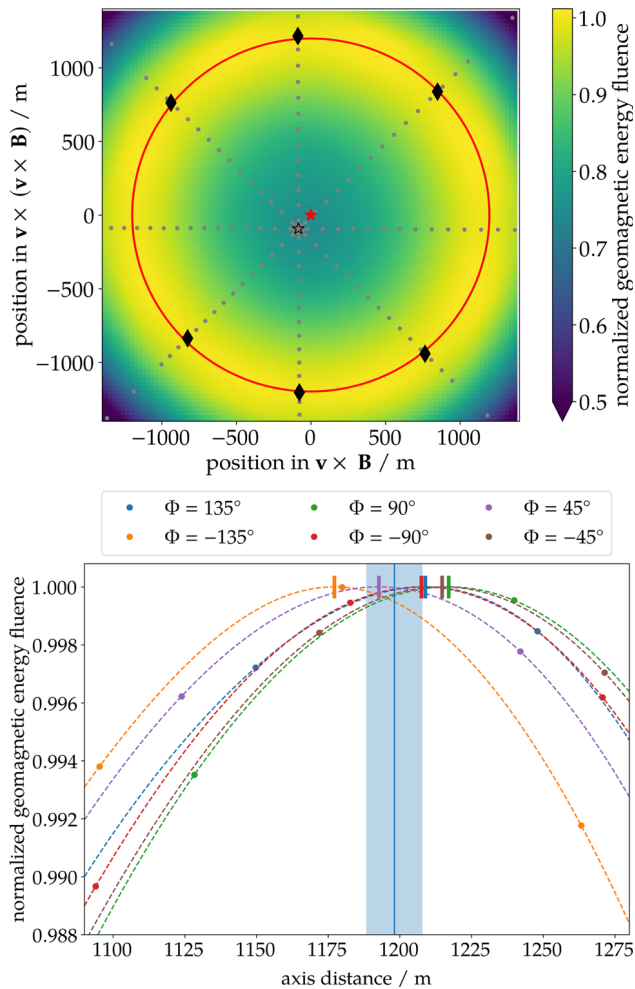

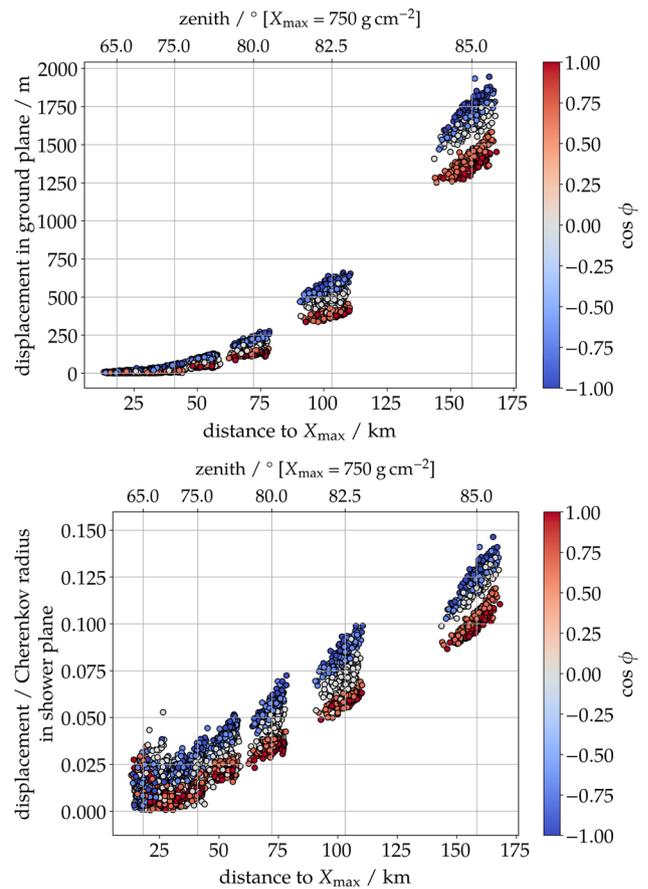

**Fig. 3** Result of the the iterative fit procedure to estimate the radio symmetry center. The geomagnetic energy fluence is normalized to the maximum along each arm. For the fit, only the position of the Cherenkov ring along the arm, and not its signal strength, is used. Top: 2D visualization of the fitted Cherenkov ring. For illustration purposes the background constitutes the cubic interpolation of the geomagnetic energy fluence from signals around the Cherenkov ring (signals on or close to the $\mathbf{v} \times \mathbf{B}$-axis are recovered using the found radio symmetry center.). In red the fitted Cherenkov ring and its center, the radio symmetry center, are shown. The black star marks the position of the MC impact point, grey dots show the positions of the simulated pulses. The positions of maximal geomagnetic energy fluence found for each arm of the star-shaped grid are denoted by the black diamonds. Bottom: 1D lateral distribution of the geomagnetic energy fluence for each polar angle of the star-shaped grid except for the $\mathbf{v} \times \mathbf{B}$ axis. Colored points denoted the calculated geomagnetic energy fluence for the simulated pulses. Their interpolation is shown by the dashed lines for each arm, the position of the maximum geomagnetic energy fluence is marked by the colored vertical line. The blue line and box denote the fitted radius of the Cherenkov ring and its uncertainty. The axis distances displayed on the x-axis are calculated using the fitted radio symmetry center

**Fig. 4** Top: displacement of the radio symmetry center with respect to the MC impact point in the ground plane as function of distance to $X_{\text{max}}$. Bottom: displacement of the radio symmetry center with respect to the MC impact point in the shower plane normalized to the fitted radius of the Cherenkov ring as function of distance to shower maximum. The color-coded cosine of the azimuth arrival direction illustrates an East ($\cos\phi = 1$) West ($\cos\phi = -1$) asymmetry

shower axis in the direction of $X_{\text{max}}$. For inclined showers the integral can only be solved numerically as the atmospheric curvature needs to be taken into account. In the first order $d_{\text{max}}$ scales with the zenith angle, and only in second order with $X_{\text{max}}$. For a displaced radio symmetry center, the geometrical distance between this point and the shower maximum is smaller. However the deviation is of the order of $\lesssim 1\%$ and therefore negligible for our purposes.

In Fig. 4 we summarize the observed displacement between MC impact point and radio symmetry center. In the ground plane we find a displacement of more than 1500 m for the highly inclined showers (top). This is of the same order as the spacing of the detector stations for the radio upgrade of the Pierre Auger Observatory [4]. To put the magnitude of the displacement into context we also express the offset in the shower plane as a fraction of the fitted radius of the Cherenkov ring which can go up to 15% (bottom). The presented displacement exhibits a pronounced scatter. The cosine of the azimuth angle, denoted by the color, shows

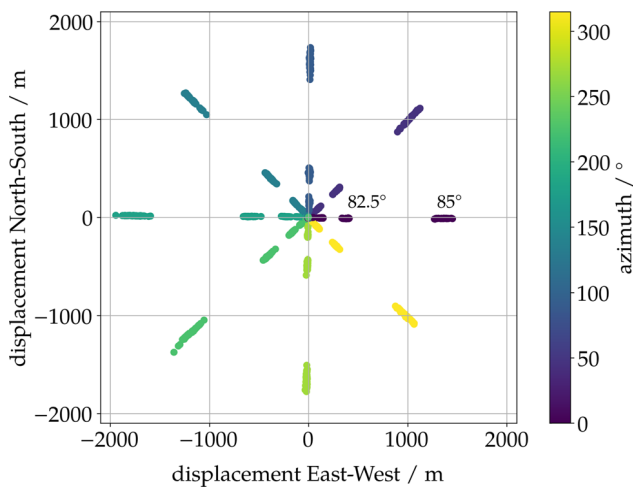

**Fig. 5** Displacement of the radio symmetry center in the ground plane relative to the MC impact point in the coordinate origin. North is defined as geomagnetic North. The radio symmetry center is always displaced into the incoming direction of the showers. Hence, the clustering of points originates from the binned MC arrival direction of our set of simulations. For the two most inclined bins their MC zenith angle is annotated in the plot

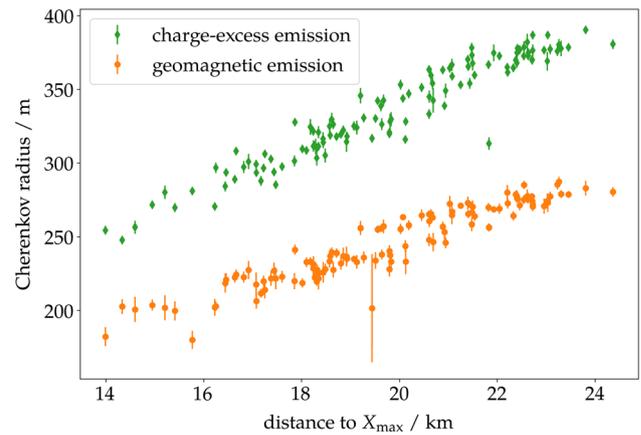

**Fig. 6** Fitted Cherenkov radius of the geomagnetic and charge-excess emission contributions individually as function of the distance to the shower maximum. Only showers with $\sin\alpha < 0.25$, i.e., coming from North are used here

### 3.3 Investigation of the Cherenkov radius in the geomagnetic and charge-excess emission for showers with a small $\sin\alpha$

For simulations with a small geomagnetic angle, i.e. $\sin\alpha < 0.25$, the charge-excess contribution is relatively strong. This allows for an independent analysis of the Cherenkov radius from the charge-excess and geomagnetic emission. As mentioned earlier, for those showers it is not possible to accurately determine the symmetry center. Therefore, the ring center is fixed to the MC impact point and just the ring radius is fitted with the method given above.

Already in Fig. 2 a difference in the Cherenkov radii found for the geomagnetic and charge-excess contributions is visible. This is confirmed for the 120 showers with $\sin\alpha < 0.25$ as shown in Fig. 6. As the center is fixed, all showers are fitted successfully and no further selection is applied. The estimated radius of the Cherenkov ring is systematically larger for the charge-excess contribution than for the geomagnetic emission. Assuming that the radio emission originates from a point source at an altitude $h$, the Cherenkov angle is given by $\theta_{Ch} = \arccos(1/n(h))$. Hence, one can estimate the height $h$ of the emission region for the given Cherenkov radius. We find that the charge-excess originates from higher up in the atmosphere than the geomagnetic emission. This is in agreement with the results found in [2]. In [25] it is reported that the charge-excess induced current peaks deeper in the atmosphere. A possible explanation for this seeming contradiction is that the coherent emission which arrives at ground does not primarily originate form the location of the highest current due to interference effects. Further studies are needed to clarify this question.

A rotational asymmetry in the charge-excess emission was reported in [2] for one example event. Here, we investigate that the found displacement depends on the shower arrival direction. The displacement is strongest for showers coming from West and weakest for East (given the inclination of the magnetic field of $\sim 36°$ both directions translate to the highest $\sin\alpha$ values). This dependency is further investigated in Fig. 5 where we show the position of the fitted radio symmetry center on ground with respect to the MC impact point in the coordinate origin. We observe a displacement in the direction from which the shower is incoming, i.e., a displacement towards the shower maximum with a small rotation. The previously described scatter manifests as an East-West asymmetry. As the atmosphere in CoREAS simulations is rotationally symmetric these asymmetries in the displacement cannot be cause by atmospheric properties. An intuitive explanation is provided by the deflection of the charged particles in the Earth's magnetic field. Given that the majority of shower particles is negatively charged, one can assume that the shower's particle bary-center is displaced from the MC axis in the direction of the Lorentz Force for a negatively charged particle. Thus, the particle bary-center for showers from west would be displaced below the shower axis, i.e., towards west, while showers from east would be displaced above the shower axis, i.e., also towards west. Hence, this additional displacement already in the particle cascade would add up with the displacement due to refraction and could cause the observed asymmetry. However, further investigations are needed to establish this cause.

the charge-excess emission of 120 showers with low sin $\alpha$ and compare the energy fluence found on the Cherenkov ring for all arms of the star-shaped grid (except of the $\mathbf{v} \times \mathbf{B}$ axis) normalized to the average energy fluence over all arms on an event-to-event basis. We find deviations from rotational symmetry with a standard deviation of 9% in energy fluence. We do not find a convincing proof of a preferred orientation of this asymmetry, but due to the low event number we can also not exclude one. A random spread could possibly be introduced by air-shower sub-structure originating in the hadronic cascade. We note that indications for a possible random deviation from rotational symmetry in the charge-excess emission of inclined air showers have been noticed before.[2]

### 3.4 Comparison of the radio symmetry center displacement for different frequency bands

So far we have analysed the radio-emission for frequencies in the 30 to 80 MHz band. This band is used by most current-generation radio experiments and in particular also by the upcoming large-scale Auger radio detector [4]. We now determine the displacement of the symmetry center for footprints in the 50 to 200 MHz frequency band. This is the target frequency band of the GRAND experiment [12], currently being in a proposal state, which is also focused on radio measurements of inclined air showers. In Fig. 7 the fitted symmetry center displacement (top) and Cherenkov radius (bottom) are shown for the two frequency bands.

We find no difference in the average behaviour of the symmetry center displacement between the two frequency bands. The spread is smaller for higher frequencies as the Cherenkov ring is more pronounced and thus easier to fit. On average the fitted Cherenkov radius is $\sim$ 5% larger for the higher frequency band. This trend is in agreement with [24] which showed the Cherenkov radius increasing with the frequency for vertical showers. This might be caused by differences in the geometrical distribution of the shower particles which primarily contribute to the radio signal in the considered frequency ranges. However, the exact origin of this deviation needs future investigations which are beyond the scope of this paper.

## 4 Interpretation of the displacement as due to refraction

We have shown that for simulations in an atmosphere with a varying refractive index the radio symmetry center is systematically displaced from the MC impact point. In this section we show that this displacement is in agreement with refraction of radio waves in a refractive atmosphere as described by Snell's law (cf. Eq. (4)). For this purpose we develop a model simulating the propagation of a single electromagnetic wave. Furthermore we summarize and validate the treatment of the refractivity in CoREAS and discuss the validity of our model.

---
[2] Private communication with F. Briechle.

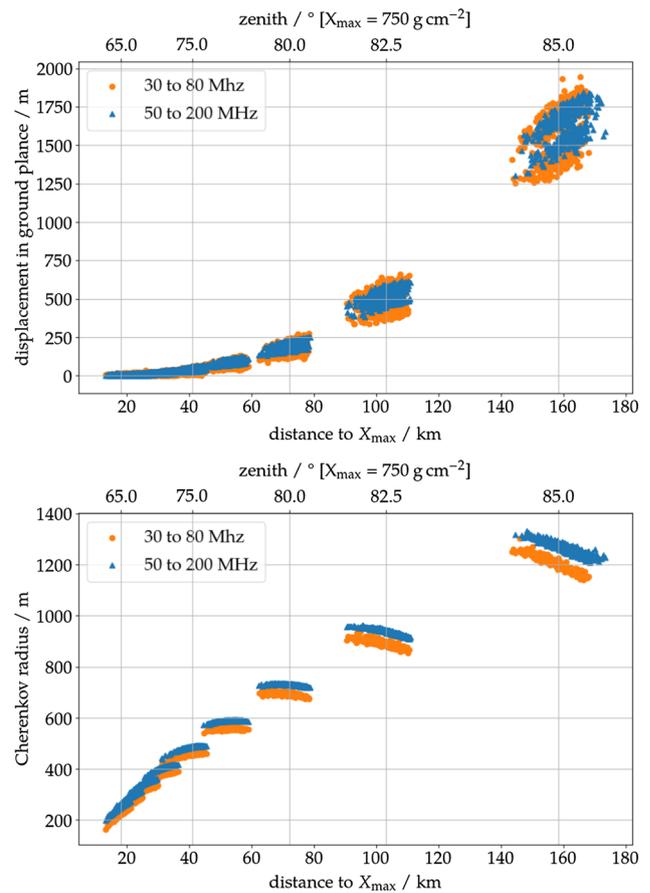

**Fig. 7** Top: displacement of the radio-emission footprint at ground for two different frequency bands. Bottom: fitted radius of the Cherenkov ring in the geomagnetic emission for the two different frequency bands

### 4.1 Description of refraction using Snell's law

We study the propagation of a single electromagnetic wave through the Earth's atmosphere described by a curved trajectory undergoing refraction according to Snell's law. For this purpose we assume discrete changes of the refractive index along the edges of imaginary layers throughout the atmosphere. The propagation within a layer with an upper edge height $h_i$ is described by a straight uniform expansion with the phase velocity $c_n = c_0/n(h_i)$ given the refractive index as function of the height above sea level $n(h)$. We adopt the refractive index as frequency-independent (i.e., non-dispersive) in the band from 30 to 200 MHz that we consider here. The change of direction between two layers with $n_1$ and $n_2$ is described in terms of the incidence angle ($\vartheta$) from $\vartheta_1$ to $\vartheta_2$ following Snell's law

$$\frac{\sin \vartheta_2}{\sin \vartheta_1} = \frac{n_1}{n_2}. \quad (4)$$

The refraction is calculated in a curved atmosphere. The relationship between the geometrical distance from ground $d_g$ and height above ground $h_g$ is given for a zenith angle $\theta$, observation level $h_{obs}$ and the Earth's radius $r_{earth} = 6371$km by

$$d_g^2 + 2(r_{earth} + h_{obs})(d_g \cos\theta - h_g) - h_g^2 = 0. \quad (5)$$

By solving this quadratic equation, one can calculate the height above sea level for every given distance to the ground by $h = h_g(d_g, \theta) + h_{obs}$. The refractivity $N \equiv n - 1$ at a given height is calculated according to the density profile of the given atmospheric model and the given refractivity at sea level ($N_0$),

$$N(h) = N_0 \frac{\rho(h)}{\rho(0)}. \quad (6)$$

We employ an atmospheric model with four exponential layers and one linear layer as used in CORSIKA/CoREAS, implementation from [26,27]. The thickness of each layer is set to 1 m assuring a high accuracy of the calculation.[3]

To predict the magnitude of a symmetry center displacement by refraction, we simulate the propagation of an electromagnetic wave along a bent trajectory with an initial direction aligned to the MC axis for a shower with a given zenith angle, towards the ground plane. The intersection between the bent trajectory and the ground plane is compared to the intersection between the ground plane and the MC axis. Given these two points, the symmetry center displacement can be inferred as depicted in Fig. 8. In Fig. 9 the predicted symmetry center displacement along the ground plane is shown as function of the geometrical distance along the MC axis for shower geometries with a zenith angle between 65° and 85°. The orange line symbolises the displacement for a source at a fixed slant depth of 750 g/cm² (e.g., shower maximum, average depth of maximum of our set of simulated showers). For a given slant depth, this distance translates to a zenith angle (top x-axis). Our model predicts a displacement of the order of 1.5 km for the most inclined showers at $\theta = 85°$. In orange squares the displacement is shown for different slant depths between 620 and 1000 g/cm² (typical range in our set of simulated showers) along the MC axis for 5 different zenith angles ($\theta = 65°, 75°, 80°, 82.5°, 85°$). The model predictions are compared to the displacements determined from the CoREAS simulation set (colored circles: cf. Sect. 3, Fig. 4). The displacement is reasonably described by our model in terms of the overall magnitude (orange line) as well as the slope as function of the source's slant depth (orange

---
[3] Since we account for the curvature of the Earth this does not equal 1 m in change of height between two layers.

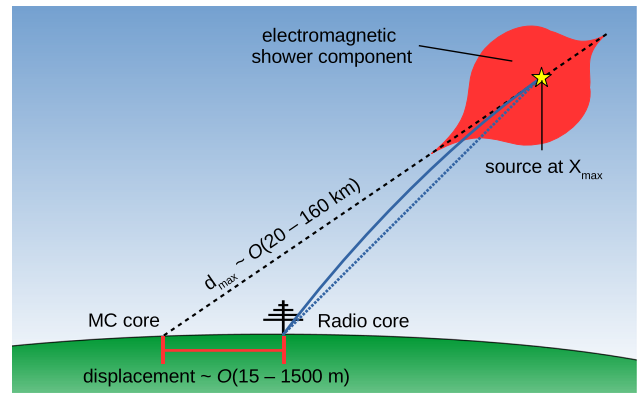

**Fig. 8** Illustration for refraction in the atmosphere. The star illustrates the source, e.g., the shower maximum. The dashed black line illustrates the MC shower axis, it's intersection with the ground plane defines the MC impact point. The solid blue line illustrates a curved trajectory with same initial direction as the MC axis, the blue dotted blue line a straight line between source and an observer. The intersection of the curved trajectory with the ground defines the radio symmetry center. The arising curvature and symmetry center displacement are over-emphasized

squares). In the bottom frame we show the absolute residuals between CoREAS displacement and refractive model. For their calculation we interpolated the model prediction along the orange squares to match the actual slant depth of the shower maximum of the simulated air showers. The residuals show no strong correlation with depth of shower maximum and increase up to ∼ 250 m for the most inclined showers. Furthermore, our model predicts a displacement always towards the shower incoming direction. This corresponds to a refraction towards the ground, i.e., decreasing angle of incidence, which is given by a radially symmetric atmosphere (cf. Fig. 8). In Fig. 5 this behaviour was also observed for CoREAS simulations as the simulated showers exhibit a radio symmetry center displacement almost entirely in the incoming direction of the shower. As emphasised earlier, an East-West asymmetry as seen in CoREAS simulations, cf. Figs. 4 and 5, cannot be described by refractivity.

We verified that the impact of the atmospheric model, i.e. the density profile, is below 3% between the US Standard Atmosphere after Keilhauer and the Malargüe October atmosphere [20]. Comparing different observer altitudes we find no difference for the displacement as function of $d_{max}$. As already shown in Fig. 1, the refractivity at sea level has an influence on the predicted displacement. The yearly fluctuations of the air refractivity at the site of the Pierre Auger Observatory amount to 7% [2]. Varying $N_0$ over a range of ±15% we find the displacement to scale linearly with $N_0$.

### 4.2 Refraction and its treatment in CoREAS

For the numerical calculation of the radio emission of an extensive air shower for an observer at ground, the refractive

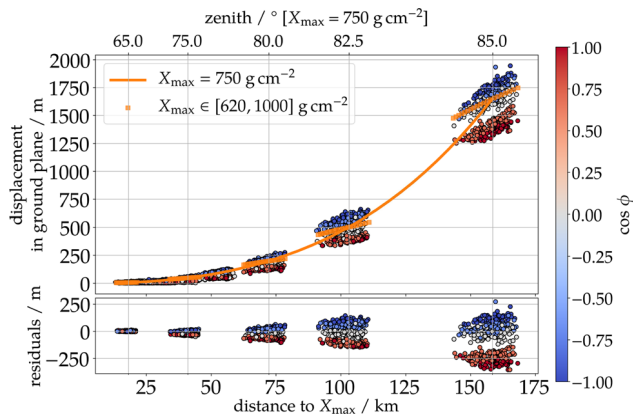

**Fig. 9** Comparison between model-predicted and CoREAS-derived displacement of the radio symmetry center. The displacement is expressed within the ground plane. The orange line constitutes our model prediction for a source at a fixed slant depth of $X_{\mathrm{max}} = 750$ g/cm$^2$ (translation to zenith angle on the top x-axis). The orange squares show the displacement as function of the source slant depth (e.g. $X_{\mathrm{max}}$) for each given zenith angle (cf. top x-axis). The colored circles show the displacement determined from the CoREAS simulation set, cf. Fig. 4. The residuals are shown in the bottom frame

index has to be taken into account for two processes: first, in the generation of the radio emission for each particle; and second, in the the propagation of each electromagnetic wave from a source to an observer. In CoREAS, the former is realistically included in the calculation of the radio emission from each particle using the endpoint formalism [17,28,29]. However, the treatment of the propagation is approximated. Since electromagnetic waves in the radio regime do not suffer from any significant attenuation effects while propagating through air, this propagation is described entirely by two quantities. First, the geometrical distance ($d$), that the radio wave passes between source and observer, as the intensity of the emission scales with this distance. And second, the light propagation time ($t_n$) between source and observer which is of crucial importance as it governs the coherence of the signal seen by an observer from the full air shower. In CoREAS, $t_n$ is calculated taking into account a refractive index dependent (phase-) velocity of the emission

$$t_n = \frac{1}{c} \int_{\mathrm{source}}^{\mathrm{observer}} n(h(\ell)) \, d\ell. \quad (7)$$

To calculate both quantities, CoREAS assumes a straight path between source and observer (cf. Fig. 8: dashed line). This approximation has implications for the geometrical distance between source and observer as a straight line underestimates the real distance along a curved trajectory. For the calculation of the light propagation time an additional implication arises from the fact that the average refractivity along a straight line varies from the refractivity along a curved trajectory. We find that the average refractivity and consequently the light propagation time is overestimated along straight trajectories.

We stress that the description of the propagation of the radio emission along straight trajectories in CoREAS is not in contradiction with the above-established refraction of the radio emission and the resulting displacement of the radio symmetry center in CoREAS simulations. In fact, the refraction of radio waves is a consequence of the fact that the propagation velocity changes with the refractive index $c_n$. It is in fact possible to achieve a displacement of the whole coherent signal pattern at ground by an accurate description of the light propagation time along straight trajectories, as we will demonstrate below.

To verify if the calculation along straight trajectories between source and observer is sufficiently accurate to calculate the radio emission seen from a full extensive air shower, we determine the geometrical distance and light propagation time following bent and straight trajectories (cf. Fig. 8) for several geometries. We simulate the propagation of an electromagnetic wave given incoming direction and atmospheric depth along a bent trajectory towards the ground plane. Once the trajectory intersects with this ground plane the process is stopped and $d_{\mathrm{curved}}$ and $t_{\mathrm{curved}}$ are calculated via a sum of $d_i$, $t_{n_i}$ over all layers. Given the intersection and the initial starting point in the atmosphere, a straight trajectory is defined and $d_{\mathrm{straight}}$ and $t_{\mathrm{straight}}$ are calculated for comparison.

In Fig. 10 (Top) the geometrical distance is compared between curved and straight trajectories in absolute terms of $d_{\mathrm{curved}} - d_{\mathrm{straight}}$, given the ambient conditions used for the above introduced simulation set. The comparison is shown as function of the geometrical distance along the straight trajectory between source and observer. The source positions are set to be at an atmospheric depth of $X = 750$ g/cm$^2$ for incoming directions with zenith angles between 65° and 85°. We obtain a maximal error of around 4 cm for the most inclined geometries with a path distance of $\sim 150$ km. With a relative deviation of less than $1 \times 10^{-6}$ this approximation is therefore completely suitable.

For the light propagation time, the relative difference for two source positions and one observer position between curved and straight trajectories $\sigma_t = \Delta t_{\mathrm{curved}} - \Delta t_{\mathrm{straight}}$ is of relevance as it governs the coherence of the total signal seen by a given observer. We do not have an analytic description for curved trajectories, however we can employ our model to determine the observer position at ground $\mathbf{O}(\mathbf{P}, \hat{\theta})$ for every given source position $\mathbf{P}$ and initial direction $\hat{\theta}$. Hence, to find two sources connected with curved trajectories to one observer at ground we have to find the initial direction $\hat{\theta}_2$ for a given second source which defines a trajectory that connects this source to an observer given by the first source and direction $\mathbf{O}_1(\mathbf{P}_1, \hat{\theta}_1)$. For this purpose we employ a root-finding algorithm that solves the following equation for $\hat{\theta}_2$: $\mathbf{O}_2(\mathbf{P}_2, \hat{\theta}_2) - \mathbf{O}_1(\mathbf{P}_1, \hat{\theta}_1) = 0$ ($\mathbf{P}_1, \mathbf{P}_2, \hat{\theta}_1$ are fixed). When the correct $\hat{\theta}_2$ is found, i.e., $\mathbf{O} = \mathbf{O}_1 = \mathbf{O}_2$, the propagation between $\mathbf{P}_1$ or $\mathbf{P}_2$ and $\mathbf{O}$ is evaluated for curved and straight trajectories and $\sigma_t$ is calculated. Figure 10 (bottom) shows $\sigma_t$ for different geometries and configurations of $\mathbf{P}_1$ and $\mathbf{P}_2$. The timing error $\sigma_t$ between two sources which are located on one axis with a zenith angle between 65° and 85° and depths of 1000 and 400 g/cm$^2$ is shown by the orange line. This range of atmospheric depths covers the bulk of the radiation energy release from the longitudinal development of an extensive air shower [2, Fig. 5]. In the most extreme case, the error in the relative arrival times for a source at the beginning of the shower evolution and a source at the end of the shower evolution, estimated using straight tracks, amounts to $\sigma_t \lesssim 0.1$ ns. This is well below the oscillation time of electromagnetic waves in the MHz regime. The blue line in the same figure demonstrates the errors made for sources laterally displaced by a shift of ±655 m above and below the shower axis along an axis perpendicular to the shower axis at a depth of 750 g/cm$^2$. This value was chosen such that it matches the Molière radius expected for a shower with $\theta = 85°$ at $X_{\max} = 750$ g/cm$^2$ [30]. The errors due to the straight-line approximation are even much smaller.

While it may seem paradoxical on first sight that a calculation approximating propagation of electromagnetic waves along straight tracks can yield refractive ray bending, we have shown that the relevant calculation of relative arrival times is described well within the needed accuracy, i.e., is fully adequate for this purpose. We note that, similar to our findings, it was already found based on analytic calculations in reference [31, cf. Fig 9] that a straight-line approximation is sufficient for the calculation of relative arrival times of radio waves in extensive air showers.

Additionally the refractive ray bending changes the incoming direction of the radio emission. This has implications for the reconstruction of the radio emission with real radio antennas as their response pattern is direction-dependent. We find a maximum change of direction of $\sim 0.14°$, which is in agreement with [32]. This is below current experimental accuracy as well as the change in the incoming direction between early and late observers on the ground plane, estimated as $\mathcal{O}(1°)$ for a 85° shower.

## 5 Conclusions

We have established that the radio symmetry center is displaced with respect to the MC impact point in CoREAS simulations of inclined air showers. This displacement shows no significant dependence on the considered frequency band for non-dispersive refractivity. We have developed a model which reproduces this displacement quantitatively, describing it as a result of refraction of the radio waves during propagation in an atmosphere with a refractive index gradient.

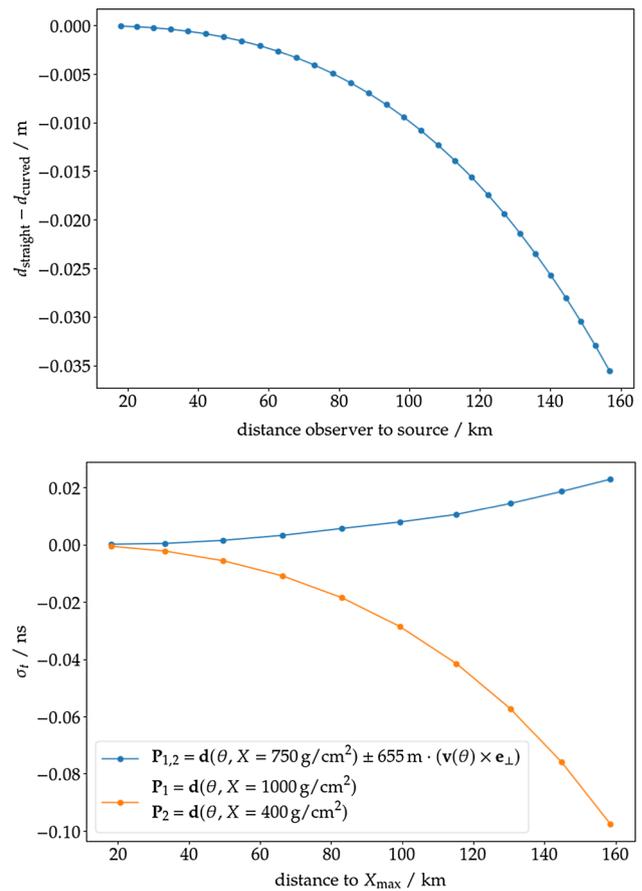

**Fig. 10** Top: difference in geometrical distance between a source at a depth of $X = 750$ g/cm$^2$ and an observer at ground level for a straight-track calculation and curved-track calculation for zenith angles from 65° to 85°. Bottom: difference in the relative arrival times calculated with curved and straight-track propagation $\sigma_t = \Delta t_{\text{curved}} - \Delta t_{\text{straight}}$ arising for two source positions and one observer position. Distance to $X_{\max}$ is calculated using a fixed depth of $X_{\max} = 750$ g/cm$^2$

We have also discussed the validity of approximations made in CoREAS and shown that they are adequate to describe refractive effects in air-shower radio simulations. We have found indications that there are secondary effects causing additional scatter in the displacement of the radio symmetry center which is not related to the refractive index of the atmosphere or the atmosphere in principle but could be caused by the geomagnetic field.

These findings have several implications towards the observation of cosmic rays with the radio detection technique. For the development of reconstruction algorithms, assuming the MC impact point as symmetry center of the radio-emission footprint will disturb its lateral distribution, causing a mismodelling of the signal distribution. In observations the refractive displacement primarily has to be taken into account in the interpretation of the reconstructed shower geometry. Considering hybrid detection and reconstruction, refractive displacement has to be taken into account when comparing/combining results across different detection techniques.

**Acknowledgements** We would like to thank our colleagues involved in radio detection within the Pierre Auger Observatory for very fruitful discussions. We are grateful to F.G. Schröder and O. Scholten for their valuable comments on our manuscript. Financial support by the BMBF Verbund-forschung Astroteilchenphysik is acknowledged. Felix Schlüter is supported by the Helmholtz International Research School for Astroparticle Physics and Enabling Technologies (HIRSAP) (Grant number HIRS-0009). Simulations for this work were performed on the supercomputer ForHLR at KIT and on the pleiades Cluster at University of Wuppertal. ForHLR is funded by the Ministry of Science, Research and the Arts Baden-Württemberg and the Federal Ministry of Education and Research. Pleiades is supported by the Deutsche Forschungsgemeinschaft (DFG).

**Data Availability Statement** This manuscript has no associated data or the data will not be deposited. [Authors' comment: The data are available from the authors upon request.]